\documentclass[twocolumns,aps,prb,reprint,superscriptaddress,showpacs]{revtex4-1}
\usepackage{graphicx}% Include figure files
\usepackage{subfigure}
\usepackage{dcolumn}% Align table columns on decimal point
\usepackage{bm}% bold math
\usepackage{hyperref}% add hypertext capabilities
\usepackage{indentfirst}
\usepackage{braket}
\usepackage{amsmath}
\usepackage{amssymb}
\hypersetup{colorlinks=true, citecolor=blue, urlcolor=blue, linkcolor=blue}
\begin{document}

\title{Quantum phase diagram and chiral spin liquid in the extended spin-$\frac{1}{2}$ honeycomb XY model}


%\thanks{}
\author{Yixuan Huang}
\affiliation{Texas Center for Superconductivity, University of Houston, Houston, Texas 77204, USA.}
\author{Xiao-Yu Dong}
\affiliation{Department of Physics and Astronomy, California State University, Northridge, California 91330, USA.}
\affiliation{Department of Physics and Astronomy, Ghent University, Krijgslaan 281, 9000 Gent, Belgium}
\author{D. N. Sheng}
\affiliation{Department of Physics and Astronomy, California State University, Northridge, California 91330, USA.}
\author{C. S. Ting}
\affiliation{Texas Center for Superconductivity, University of Houston, Houston, Texas 77204, USA.}


\date{\today}

\begin{abstract}
The frustrated XY model on the honeycomb lattice has drawn lots of attentions because of the potential emergence of chiral spin liquid (CSL) with the increasing of frustrations or competing interactions. In this work, we study the extended spin-$\frac{1}{2}$ XY model with nearest-neighbor ($J_1$), and next-nearest-neighbor ($J_2$) interactions in the presence of a three-spins chiral ($J_{\chi}$) term using density matrix renormalization group methods. We obtain a quantum phase diagram with both conventionally ordered and topologically ordered phases. In particular, the long-sought Kalmeyer-Laughlin CSL is shown to emerge under a small $J_{\chi}$ perturbation due to the interplay of the magnetic frustration and chiral interactions. The CSL, which is a non-magnetic phase, is identified by the scalar chiral order, the finite spin gap on a torus, and the chiral entanglement spectrum described by chiral $SU(2)_{1}$ conformal field theory.
\end{abstract}

%\keywords{}

\maketitle

\textit{Introduction.}
A spin liquid~\cite{balents2010spin} features a highly frustrated phase with long range ground state entanglement~\cite{levin2006detecting,kitaev2006topological} and fractionalized quasi-particle excitations~\cite{senthil2002microscopic,PhysRevB.65.224412,sheng2005numerical} in the absence of conventional order. The exotic properties~\cite{PhysRevLett.86.1881,PhysRevLett.99.097202,isakov2011topological} of the spin liquid are relevant to both unconventional superconductivity~\cite{anderson1987resonating,rokhsar1988superconductivity,lee2006doping,wang2018chern} and topological quantum computation~\cite{nayak2008non}. Among various kinds of spin liquids, the chiral spin liquids (CSL), which have gapped bulk and gapless chiral edge excitations, is proposed by Kalmeyer and Laughlin~\cite{kalmeyer1987equivalence}. It has a nontrivial topological order, and belongs to the same topological class as the fractional quantum Hall states.

In recent years, there have been extensive studies to identify the CSL in realistic spin models on different geometries such as Kagome~\cite{gong2014emergent,bauer2014chiral,wietek2015nature,he2014chiral,zhu2015chiral}, triangle~\cite{nataf2016chiral,wietek2017chiral,PhysRevB.96.075116,PhysRevB.100.241111}, square~\cite{nielsen2013local,PhysRevB.98.184409}, and honeycomb lattices~\cite{hickey2016haldane}. Interestingly, for the XY model on honeycomb lattice theoretical studies have suggested the existence of a CSL in the highly frustrated regime that is generated by the staggered Chern-Simons flux with nontrivial topology~\cite{sedrakyan2015spontaneous,wang2018chern}. However, so far there is no direct numerical evidence supporting this claim~\cite{varney2011kaleidoscope,carrasquilla2013nature,zhu2013unexpected,di2014spiral,li2014phase,zhu2014quantum,oitmaa2014phase,bishop2014frustrated}, leaving the possible existence of a CSL in the honeycomb XY model as an open question.

Aside from the possible CSL, the XY model itself is expected to have a rich phase diagram because of the frustration induced by the next-nearest-neighbor coupling $J_{2}$. As the reminiscent of the debated intermediate phase in numerical studies, density matrix renormalization group (DMRG)~\cite{zhu2013unexpected,zhu2014quantum} and coupled cluster method~\cite{bishop2014frustrated} studies suggest an Ising antiferromagnetic state. However, exact diagonalization (ED)~\cite{varney2011kaleidoscope,varney2012quantum} and quantum Monte Carlo method studies~\cite{carrasquilla2013nature,nakafuji2017phase}
suggest a Bose-metal phase with spinon Fermi surface~\cite{sheng2009spin}. A very recent numerical study using ED reveals an emergent chiral order but the phase remains a topologically trivial chiral spin state~\cite{plekhanov2018emergent}. Up to now, the theoretical understanding of the phase diagram for honeycomb XY model is far from clear.

\begin{figure}
\centering
\includegraphics[width=0.8\columnwidth]{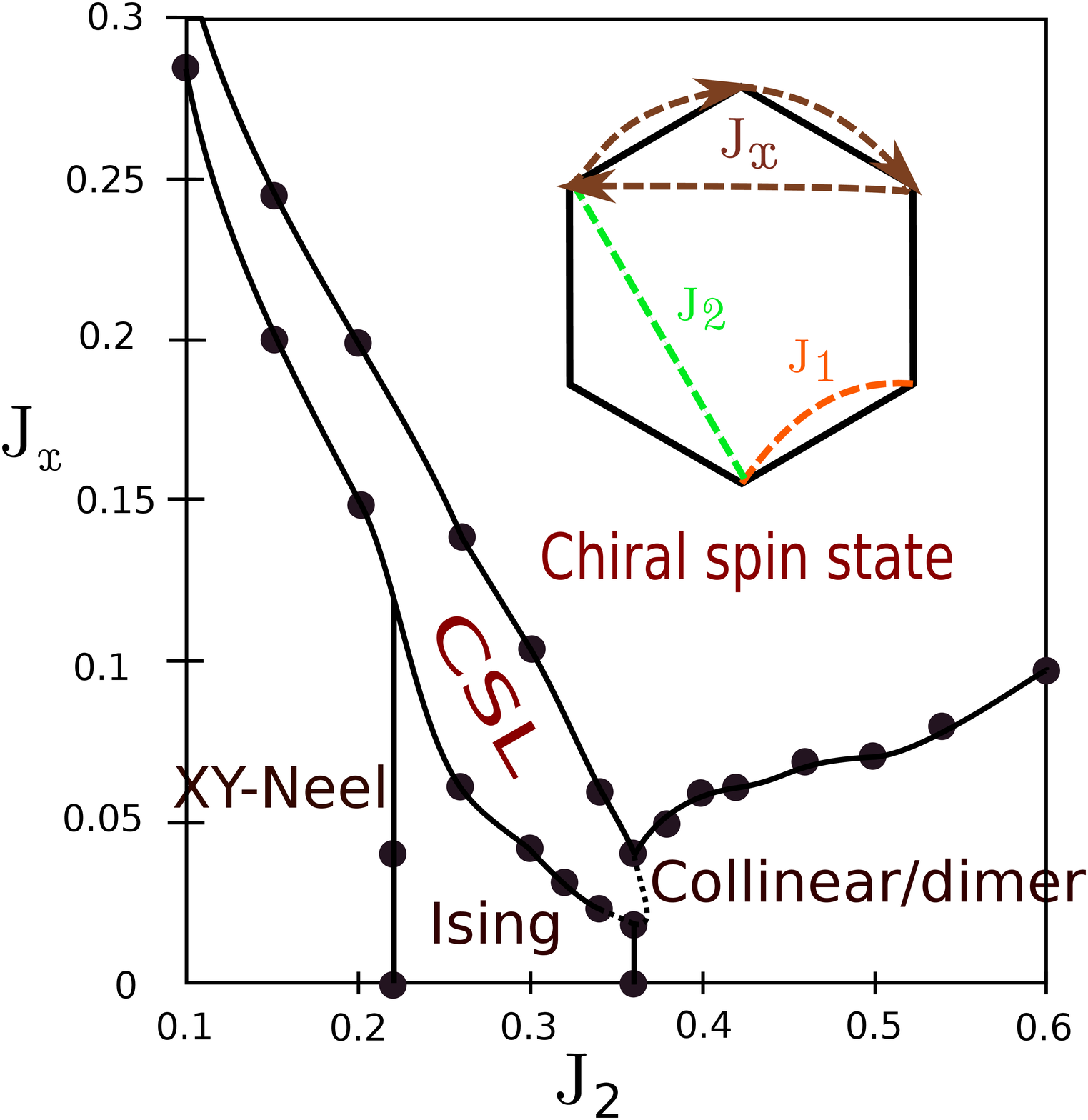}
\caption{\label{Fig1}(Color online) The schematic phase diagram of the extended XY model for $0.1<J_{2}<0.6$ and $0<J_{\chi }<0.3$, based on the results from cylindrical circumference of 4 unit cells. The CSL is identified in the intermediate regime.}
\end{figure}

The aim of this letter is to provide strong numerical evidence of the long-sought CSL in the extended spin-$\frac{1}{2}$ XY model on the honeycomb lattice and clarify the conditions for such a phase to emerge. Based on large scale DMRG~\cite{white1992density,white1993density} studies, we identify the quantum phase diagram in the presence of the nearest, next-nearest XY spin couplings and three-spins chiral interactions $\overrightarrow{S}_{i}\cdot (%
\overrightarrow{S}_{j}\times \overrightarrow{S}_{k})$. 
While there are only magnetic ordered phases in the absence of the chiral couplings, the CSL emerges with finite chiral interactions, where the minimum $J_{\chi}$ required for the emergence of the CSL appears in the intermediate $J_2$ regime. This suggests possible multiple critical points in the phase diagram, neighboring between Ising antiferromagnetic order, collinear/dimer order, and the CSL.

The CSL is identified in the extended regime above the XY-Neel state and the Ising antiferromagnetic state induced by chiral interactions. We also obtain a chiral spin state at large $J_{\chi}$ with finite chiral order. The chiral spin state shows peaks in the spin structure factor that increase with system sizes, indicating a magnetic ordered state. The phases we find without the chiral term agree with previous numerical studies using DMRG~\cite{zhu2013unexpected}. Our results demonstrate the importance of the interplay between the frustration and chiral interactions, which leads to a rich phase diagram.

%\section{Model and method}
\textit{Model and method.}
We investigate the extended spin-$\frac{1}{2}$ XY model with a uniform scalar chiral term using both infinite and finite size DMRG methods~\cite{ITensorandTenPy,tenpy} in the language of matrix product states~\cite{schollwock2011density}. We use the cylindrical geometry with circumference up to 6 (8) unit cells in the finite (infinite) size systems except for the calculations of spin gap, which is based on smaller size tori to reduce boundary effect.

The Hamiltonian of the model is given as
\begin{equation}
\label{eq1}
\begin{split}
H=J_{1}\sum\limits_{\left\langle i,j\right\rangle
}(S_{i}^{+}S_{j}^{-}+h.c.)+J_{2}\sum\limits_{\left\langle \left\langle
i,j\right\rangle \right\rangle }(S_{i}^{+}S_{j}^{-}+h.c.)\\ +J_{\chi
}\sum\limits_{i,j,k\in \triangle }\overrightarrow{S}_{i}\cdot (%
\overrightarrow{S}_{j}\times \overrightarrow{S}_{k})
\end{split}
\end{equation} 
here $\left\langle i,j\right\rangle$ refers to the nearest-neighbor sites and $\left\langle \left\langle i,j\right\rangle \right\rangle $ refers to the next-nearest-neighbor sites. $\left \{ i,j,k \right \}$ in the summation $\sum _{\Delta }$ refers to the three neighboring sites of the smallest triangle taken clockwise as shown in Fig.\ref{Fig1}. The chiral term could be derived as an effective Hamiltonian of the extended Hubbard model with an additional $\Phi $ flux through each elementary honeycomb~\cite{bauer2014chiral,hickey2016haldane,motrunich2006orbital,sen1995large}.    
We set $J_{1}=1$ as the unit for the energy scale, and use the spin U(1) symmetry for better convergence.
%Here we determine the phase diagram with XC8 and found the collinear order as the ground state.

\begin{figure}
\centering
\input{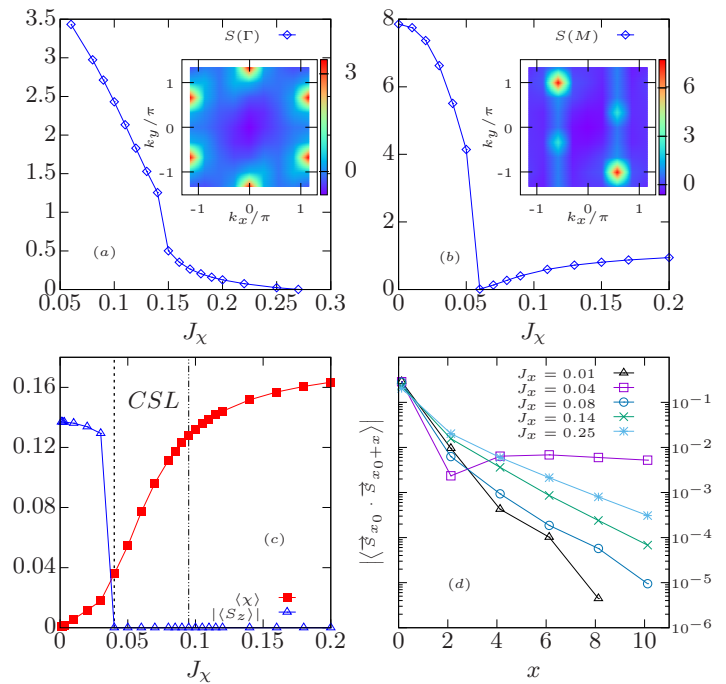}
\caption{\label{Fig2}(Color online) $(a)$ shows the peak value at $\Gamma $ point in the spin structure $ S\left (q \right ) $ at $J_{2}=0.2$ for various $J_{\chi }$ where the peak vanishes at $J_{\chi }\approx 0.15$. The inset of $(a)$ is the spin structure of the XY-Neel order at $J_{2}=0.2,J_{\chi }=0.06$, where there are clear peaks at the $\Gamma $ points in the second Brillouin zone. $(b)$ is the $M$ point peak value at $J_{2}=0.4$ for various $J_{\chi }$. The peak shows a sudden drop at $J_{\chi }\approx 0.06$, indicating a phase transition. The inset of $(b)$ is the spin structure of the collinear order at $J_{2}=0.4,J_{\chi }=0.01$, where the dominant peak is located at the $M$ points in the second Brillouin zone. $(c)$ shows the antiferromagnetic order (blue line) and the scalar chiral order (red line) at $J_{2}=0.3$ for various $J_{\chi }$ where the three corresponding phases from left to right are Ising antiferromagnetic state, CSL, and chiral spin state. The left dash line is determined by the sudden drop of antiferromagnetic order, while the right dash line is determined by the vanish of quasi-degenerate pattern in the entanglement spectrum. $(d)$ refers to the spin correlations at $J_{2}=0.3$ for various $J_{\chi }$ representing different phases. The phases at $J_{\chi }=0.01$, $0.04$, $0.08$, and $0.14$($0.25$) refer to Ising antiferromagnetic state, phase boundary, CSL, and chiral spin state respectively. The $ x_{0}$ is chosen away from the open boundary, and $x$ refers to the horizontal distance between the two spins. All of the correlations except $J_{\chi }=0.04$ show a straight line in the log plot that indicates an exponential decay. The plots above are based on finite DMRG results with $L_{y}=4\times 2$.}
\end{figure}

%\section{Phase diagram}
\textit{Phase diagram.}
The ground state phase diagram is illustrated in Fig.\ref{Fig1}.
We use spin structure factors to identify magnetic ordered phases, and entanglement spectrum to identify the topological ordered CSL. For larger $J_{\chi }$, a magnetic ordered chiral spin state with nonzero scalar chiral order is also identified.

The static spin structure in the Brillouin zone is defined as 

\begin{equation}
\label{eq2}
S\left ( \overrightarrow{q} \right ) = \frac{1}{N}\sum_{i,j}\left \langle \overrightarrow{S}_{i}\cdot \overrightarrow{S}_{j} \right \rangle e^{i\overrightarrow{q}\cdot \left ( \overrightarrow{r}_{i} -  \overrightarrow{r}_{j} \right ) }
\end{equation}

For the XY-Neel state there are peaks at the Brillouin zone $\Gamma $ points in the static spin structure as shown in the inset of Fig.\ref{Fig2}(a). The magnitude of the peak is plotted as a function of $J_{\chi }$ in Fig.\ref{Fig2}(a). It decreases rapidly as $J_{\chi }$ increases, and disappears as the system transits into the CSL at $J_{\chi }\approx 0.15$. Similarly, the peak for the collinear order at various $J_{\chi }$ is given in Fig.\ref{Fig2}(b). The inset of Fig.\ref{Fig2}(b) shows the spin structure at $J_{\chi }=0.01$ where the phase is dominant by the collinear order. The phase boundary could be identified by the sudden drop and the disappearance of the peak at $J_{\chi }\approx 0.06$. In the intermediate regime at $J_{2}=0.3$ and small $J_{\chi }$, the staggered on-site magnetization serves as the order parameter as shown in Fig.\ref{Fig2}(c). This quantity shows a sudden drop from the Ising antiferromagnetic state to the CSL at $J_{\chi }\approx 0.04$, which determines the phase boundary. The finite size analysis of it indicates a possible first order phase transition for $J_{2}$ close to 0.34, and a higher order transition for smaller $J_{2}$ (see supplemental material~\cite{SuppMaterial}).

Besides the magnetic order parameters, other properties such as the spin correlation, the entanglement entropy and spectrum are also used to identify the phase boundary. We have found consistence in those different measurements. As shown in Fig.\ref{Fig2}(d), the spin correlations are strongly enhanced at $J_{\chi }\approx 0.04$ near the phase boundary between the Ising antiferromagnetic phase and the CSL, while both phases have exponentially decaying spin correlations. The phase boundary determined by the spin correlation is the same as the one by the staggered magnetization.

Both CSL and the chiral spin state in the larger $J_{\chi }$ regime have a finite scalar chiral order that is defined as

\begin{equation}
\label{eq3}
 \left \langle \chi  \right \rangle = \frac{1}{3N}\sum\limits_{i,j,k\in \triangle }\overrightarrow{S}_{i}\cdot (\overrightarrow{S}_{j}\times \overrightarrow{S}_{k})
\end{equation}

As shown by the red curve in Fig.\ref{Fig2}(c), the chiral order increases monotonically with the increase of $J_{\chi }$ in the CSL and chiral spin state, and saturates around $ \left \langle \chi  \right \rangle \approx 0.177$. The spin correlations in these two states are given in Fig.\ref{Fig2}(d) as examples at $J_{\chi }=0.08$, and $0.14$($0.25$) respectively, where they remain exponentially decay. However, the spin correlation increases generally as $J_{\chi }$ increases. As shown in Fig.\ref{Fig4}(b), for the parameters we labeled as chiral spin state, the spin structure factors show sharp peaks, with the magnitudes of the peak values increasing with system sizes, suggesting a magnetic ordered state in the larger $J_{\chi }$ regime. We also notice that the spin structure in this chiral spin state shares the same peaks as the tetrahedral phase~\cite{hickey2016haldane,hickey2017emergence} (see supplemental material~\cite{SuppMaterial}), and we do not rule out the possibility of tetrahedral magnetic order in this regime.

%Meanwhile, the newly proposed gapless CSL~\cite{gong2019gapless} in triangle lattice also features finite chiral order as well as decaying spin correlations, and the chiral spin state would be a suitable starting point to search for such kind of state.

The extended regime of $J_{2}>0.6$ and $J_{2}<0.1$ are not our main focus in this letter because we are interested in the intermediate $J_2$ regime with strong frustration, but we do find that the CSL extends to a relatively large $J_{\chi }\approx 0.5$ at $J_{2}=0$. This implies that the CSL could survive even without the frustration induced by second nearest neighbor interactions in the XY model, which may be interesting for future study. In the regime labeled as collinear/dimer, we also find a non-magnetic dimer ground state in close competition with the collinear state at $J_{\chi }> 0.55$. As pointed out in Ref.~\cite{zhu2013unexpected}, the actual ground state depends on the system size and XC/YC geometry, and we will not try to resolve this close competition here.

The phase near the critical point of $J_{2}\approx 0.36$, $J_{\chi }\approx 0.02$ is hard to define numerically because different spin orders are mixed together in the low energy spectrum, thus the spin correlation is generally large. Here the phase boundary is measured by the unique properties of the CSL through the entanglement spectrum as discussed below, and it will be marked by the dash line as a guide to the eye.

\begin{figure}
\centering
\input{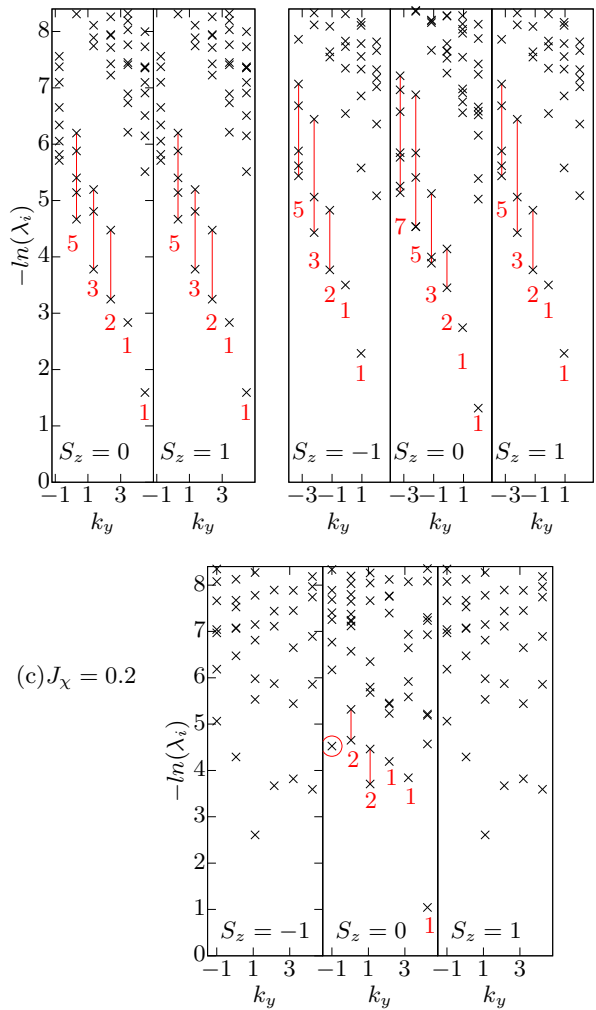}
\caption{\label{Fig3}(Color online) The ES for the spinon ground state (a) and the vacuum ground state (b) in the CSL phase at $ J_{2}=0.26, J_{\chi }=0.09$, and the ES in the chiral spin state (c) at $ J_{2}=0.26, J_{\chi }=0.2$ with different spin sectors. The spectrum is calculated using infinite DMRG with $L_{y}=6\times 2$. The $\lambda _{i}$ refers to the eigenvalues of the reduced density matrix, and the $k_{y}$ has an increasement of $\frac{2\pi }{L_{y}}$. The quasi degenerate eigenvalues are labeled by the number below each momentum. Each spin sector is separated with the help of total $S_{z}$ conservation implemented in the algorithm.}
\end{figure}

%\section{Chiral spin liquids}
\textit{Chiral spin liquids.}
The CSL is characterized by the twofold topological degenerate ground states, which are called ground state in vacuum and spinon sectors~\cite{gong2014emergent,he2014obtaining}, respectively. The entanglement spectrum (ES) of the ground state corresponds to the physical edge spectrum that is created by cutting the system in half~\cite{qi2012general,PhysRevB.86.125441,PhysRevLett.110.067208}. Following the chiral $SU(2)_{1}$ conformal field theory~\cite{francesco2012conformal}, the leading ES of a gapped CSL has the degeneracy pattern of {1,1,2,3,5...}~\cite{wen1990chiral}. As shown in Fig.\ref{Fig3}(a) and (b), the ES in the CSL phase has such quasi-degenerate pattern with decreasing momentum in the y-direction for each spin sector, though higher degeneracy levels  may not be observed due to the finite numbers of momentum sectors. The ES of the spinon ground state has a symmetry about $S_{z}=\frac{1}{2}$ which corresponds to a spinon at the edge of the cylinder, while the one of the vacuum ground state has a symmetry about $S_{z}=0$. The ES is robust in the bulk part of the CSL phase for various parameters and system sizes, but as we approach the phase boundary, additional eigenstates may also mix in the spectrum (see supplemental material~\cite{SuppMaterial}).

The main difference between the CSL and the chiral spin state is the topological edge state that can be identified through the ES. An example of the ES in the chiral spin state is also given in Fig.\ref{Fig3}(c), where the quasi-degenerate pattern disappears and additional low-lying states emerge, as opposed to the ES in the CSL shown in Fig.\ref{Fig3}(a) and (b). The phase boundary between these two states are determined mainly by the ES.

The finite chiral order represents the time reversal symmetry breaking chiral current in each small triangle, which is shown in Fig.\ref{Fig2}(c). The chiral order is significantly enhanced as the system undergoes a phase transition from the Ising antiferromagnetic state to the CSL. However, the spin correlation remains following exponential decay, as shown by the line of $J_{\chi }=0.08$ in Fig.\ref{Fig2}$(d)$. We further confirm the vanish of any conventional spin order in the CSL by obtaining the spin structure in Fig.\ref{Fig4}$(a)$, and comparing it with the one in the chiral spin state in Fig.\ref{Fig4}$(b)$. There is no significant peak in the CSL phase as opposed to other magnetic phases.

\begin{figure}
\centering
\input{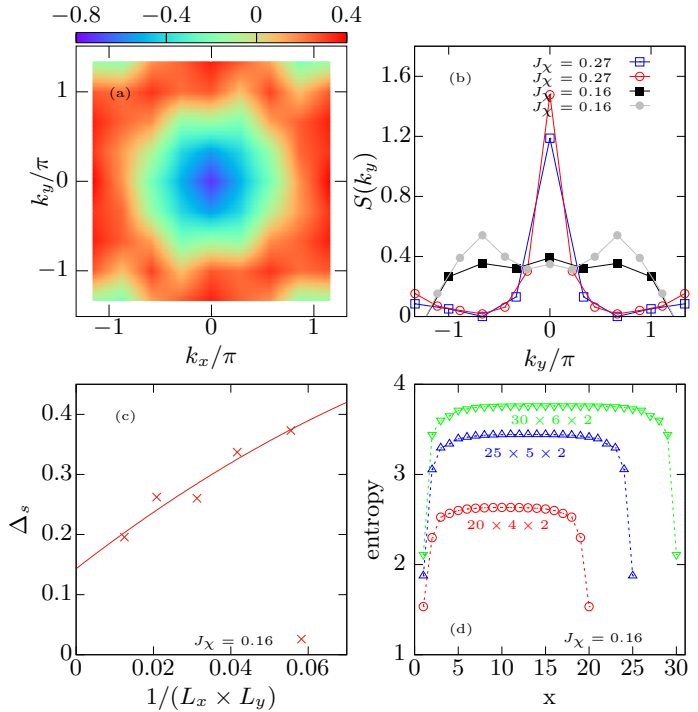}
\caption{\label{Fig4}(Color online) $(a)$ refers to the spin structure in the CSL phase at $ J_{2}=0.2, J_{\chi }=0.16$ where there is no peak as opposed to other magnetic phases. This result is based on the cluster of $20\times 4\times 2$. $(b)$ refers to the spin structure peaks with fixed $k_{x}=-\frac{2\pi }{\sqrt{3}}$ of various parameters. The blue and red lines are obtained in the chiral spin state with clusters of $20\times 4\times 2$ and $30\times 6\times 2$, respectively. The magnitude of the peak increases as the cluster size increases. The black and grey lines are obtained in the CSL with the same two clusters, where there is no significant peak. $(c)$ refers to the finite size scaling of the spin gap on the torus geometry with clusters of $3\times 3\times 2$, $4\times 3\times 2$, $4\times 4\times 2$, $6\times 4\times 2$, and $8\times 5\times 2$. $(d)$ refers to the entanglement entropy with various clusters on finite cylinders in the CSL phase. The $x$ here denotes the distance of the cut in the x direction. All of the results are obtained at $J_{2}=0.2$.}
\end{figure}

In order to identify the excitation properties of the CSL, we obtain
the spin-1 excitation gap by the energy difference between the lowest state in $S=0$ and $1$ sector. To measure the bulk excitation gap, we use the torus geometry to reduce the boundary effect. The finite size scaling of the spin gap using rectangle like clusters is shown in Fig.\ref{Fig4}$(c)$. The spin gap decays slowly as the cluster grows, and remains finite after the extrapolation, suggesting a gapped phase in the thermodynamic limit. In addition, we study the entanglement entropy of the subsystems by cutting at different bonds. As shown in Fig.\ref{Fig4}$(d)$, the entropy becomes flat away from the boundary, which corresponds to a zero central charge in the conformal field theory interpretation~\cite{calabrese2004entanglement}. This supports a gapped CSL phase that is consistent with the finite spin gap.

%\section{Summary}
\textit{Summary and discussions.}
Using large scale DMRG, we identify the long-sought CSL with the perturbation of three-spins chiral interactions in the spin-$\frac{1}{2}$ XY model on the honeycomb lattice. The CSL extends to the intermediate regime with a small $J_{\chi }$, providing evidence of the important interplay between frustration and chiral interactions driving the CSL. Here, we demonstrate that the chiral interactions are essential for the emergence of the CSL, because the minimum  critical $J_{\chi }$ of the phase transition is around $0.02$, which is stable against the increasing of system sizes, and below the critical $J_{\chi }$ there is no such quasi-degenerate pattern in the ES (see supplemental material~\cite{SuppMaterial}).

A chiral spin state is also obtained at larger $J_{\chi }$, which extends to the wider regime of $J_{2}$. The chiral spin state has a peak value for spin structure factor growing with system sizes. Further studies include finding the exact nature of this chiral spin state, and the nature of the phase transition into the CSL.

Experimentally, of all the honeycomb materials that show a quantum-spin-liquid-like behavior~\cite{nakatsuji2012spin,PhysRevLett.107.197204,PhysRevB.93.214432}, the Co-based compounds are mostly studied in the context of XY model such as $BaCo_{2}\left ( PO_{4} \right )_{2}$~\cite{nair2018short,zhong2018field} and $BaCo_{2}\left ( AsO_{4} \right )_{2}$~\cite{zhong2019weak}, thus it would be extremely interesting to search for the quantum spin liquid in such systems. On the other hand, the results of CSL may be tested in cold atoms experiments~\cite{goldman2016topological,aidelsburger2015measuring} as the spin XY model could be mapped by the bosonic Kane-Mele model in the Mott regime~\cite{plekhanov2018emergent,kane2005quantum}.

%\subsection{}
%\subsubsection{}

%\appendix
%\section{}

\textit{Acknowledgments.}
Y.H and C.S.T was supported by the Texas Center for Superconductivity and the Robert A. Welch Foundation Grant No. E-1146. Work at CSUN was supported by National Science Foundation Grants PREM DMR-1828019. Numerical calculations was completed in part with resources provided by the Center for Advanced Computing and Data Science at the University of Houston. 

\onecolumngrid
\appendix
\section{The convergence of DMRG results}

\begin{figure}
\centering
\input{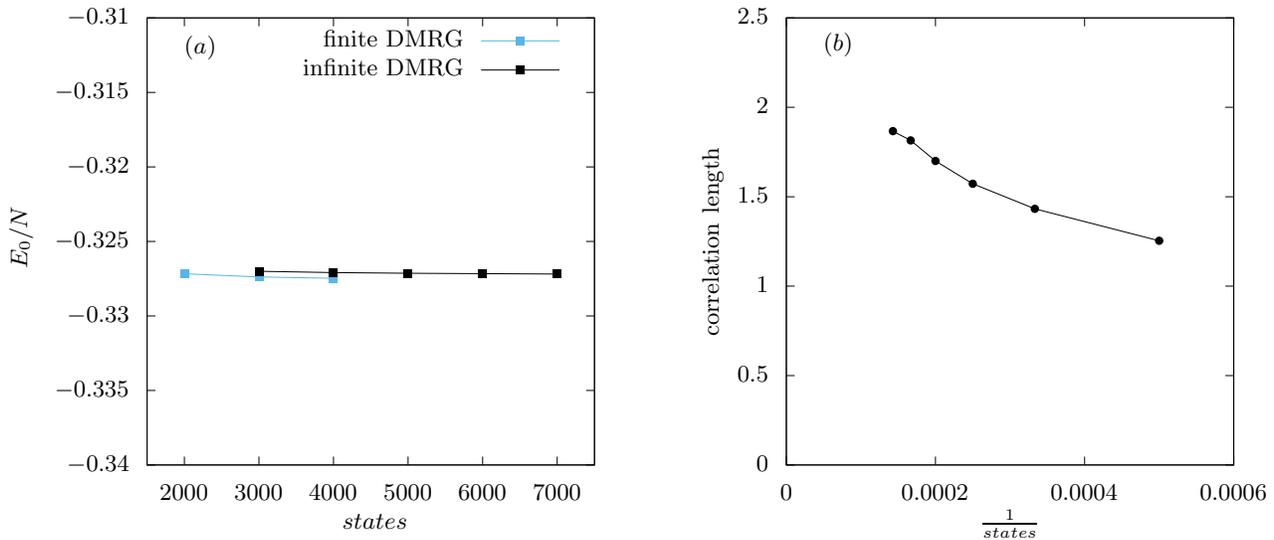}
\caption{\label{FigS1}(Color online) The ground state energy per site (a) and the scaling of the correlation length (b) with different number of states kept. The finite DMRG results are calculated with $L_{x}\times L_{y}=30\times 6\times 2$, and the infinite DMRG results are calculated with $L_{y}=6\times 2$. In the limit of infinite states kept the correlation lenth is around 2. Both results above are obtained in the CSL at $J_{2}=0.2, J_{\chi }=0.16$}
\end{figure}

\begin{figure}
\centering
\input{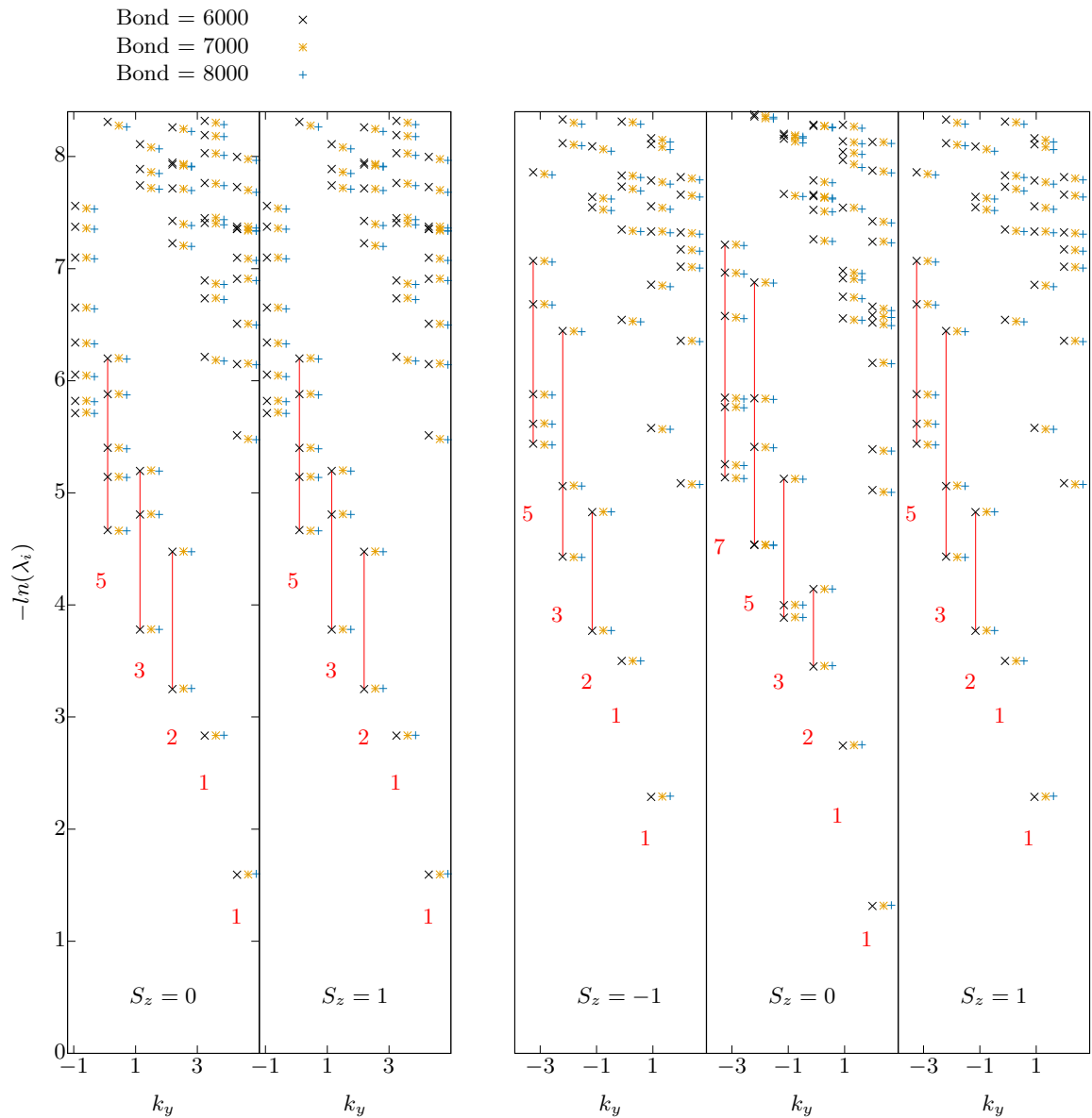}
\caption{\label{FigS2}(Color online) The entanglement spectrum of the vacuum and spinon ground state with different number of states kept. The results are calculated with $L_{y}=6\times 2$ in the CSL at $J_{2}=0.26, J_{\chi }=0.09$, in support of Fig.3 in the letter.}
\end{figure}

The convergence of Density Matrix Renormalization Group (DMRG) results could be checked by the truncation error and other physical quantities such as ground state energy, spin correlations, and entanglement entropy, with increasing number of states kept. Here the quantities with various number of states kept in the chiral spin liquid (CSL) phase is plotted in Fig.\ref{FigS1}. As shown in Fig.\ref{FigS1}(a), for both finite and infinite DMRG methods the ground state energy per site remains almost unchanged as the states increase, indicating that the results are converged. For finite DMRG method the energy per site is obtained through the center half lattices to minimize the boundary effect, and the energy per site obtained by the finite and infinite DMRG methods are very close. The entanglement spectrum also remains almost unchanged with increasing states as shown in Fig.\ref{FigS2}. We keep 3000 (6000) states for finite (infinite) DMRG for most of the calculations and are able to reach the truncation error less than $10^{-7}$ ($10^{-5}$).

\section{Finite size analysis of the phase diagram}

As shown in Fig.\ref{FigS1}(b), the correlation length in the CSL phase is around 2 after the extrapolation to infinite states kept in the numerical calculation which represents zero truncation error. Thus we believe that CSL phase found in the finite size cylinder of $L_{y}= 6\times 2$, which is larger than 2, should be valid in the thermodynamic limit.

\begin{figure}
\centering
\input{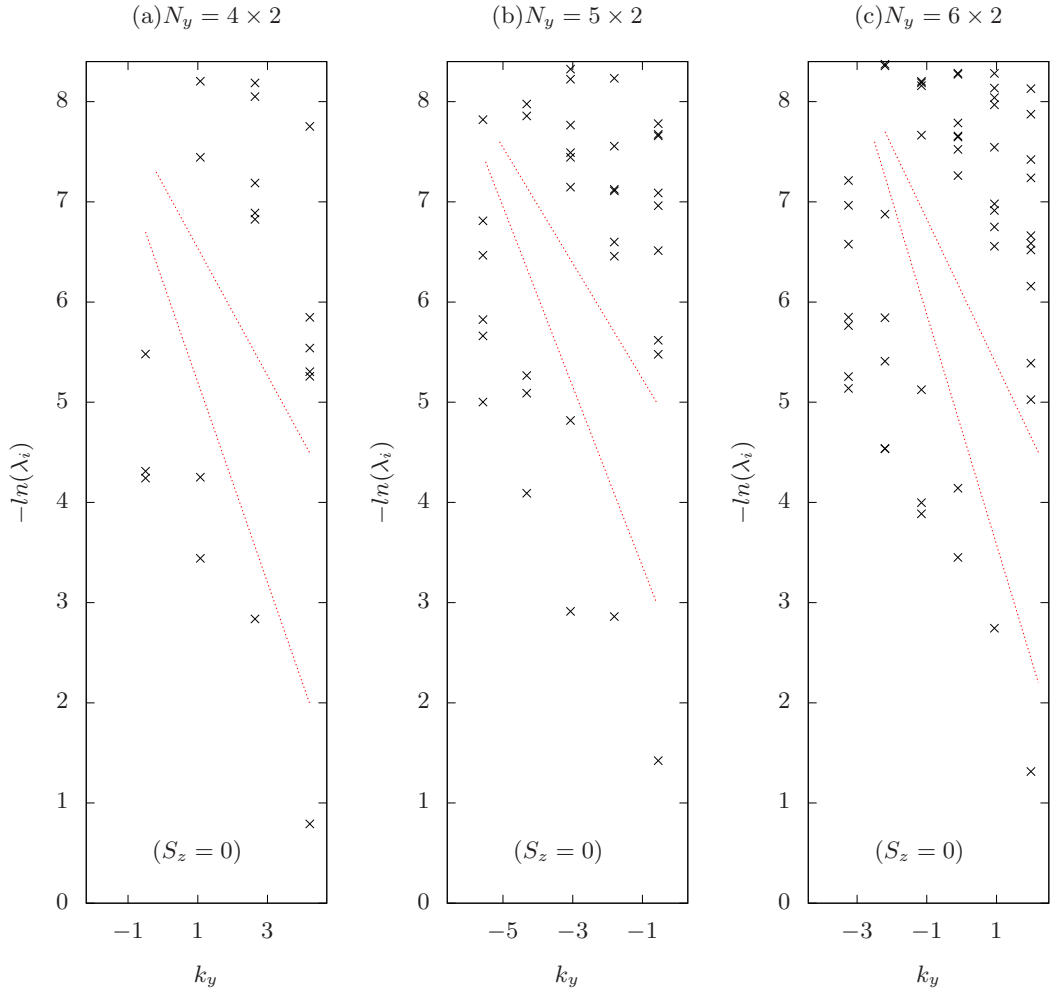}
\caption{\label{FigS10}(Color online) The entanglement spectrum for $J_{2} = 0.26, J_{\chi }=0.09$ obtained at various sysetm size, showing robust quasi-degenerate pattern in the CSL state. The dash red lines are guide to the eye.}
\end{figure}

\begin{figure}
\centering
\input{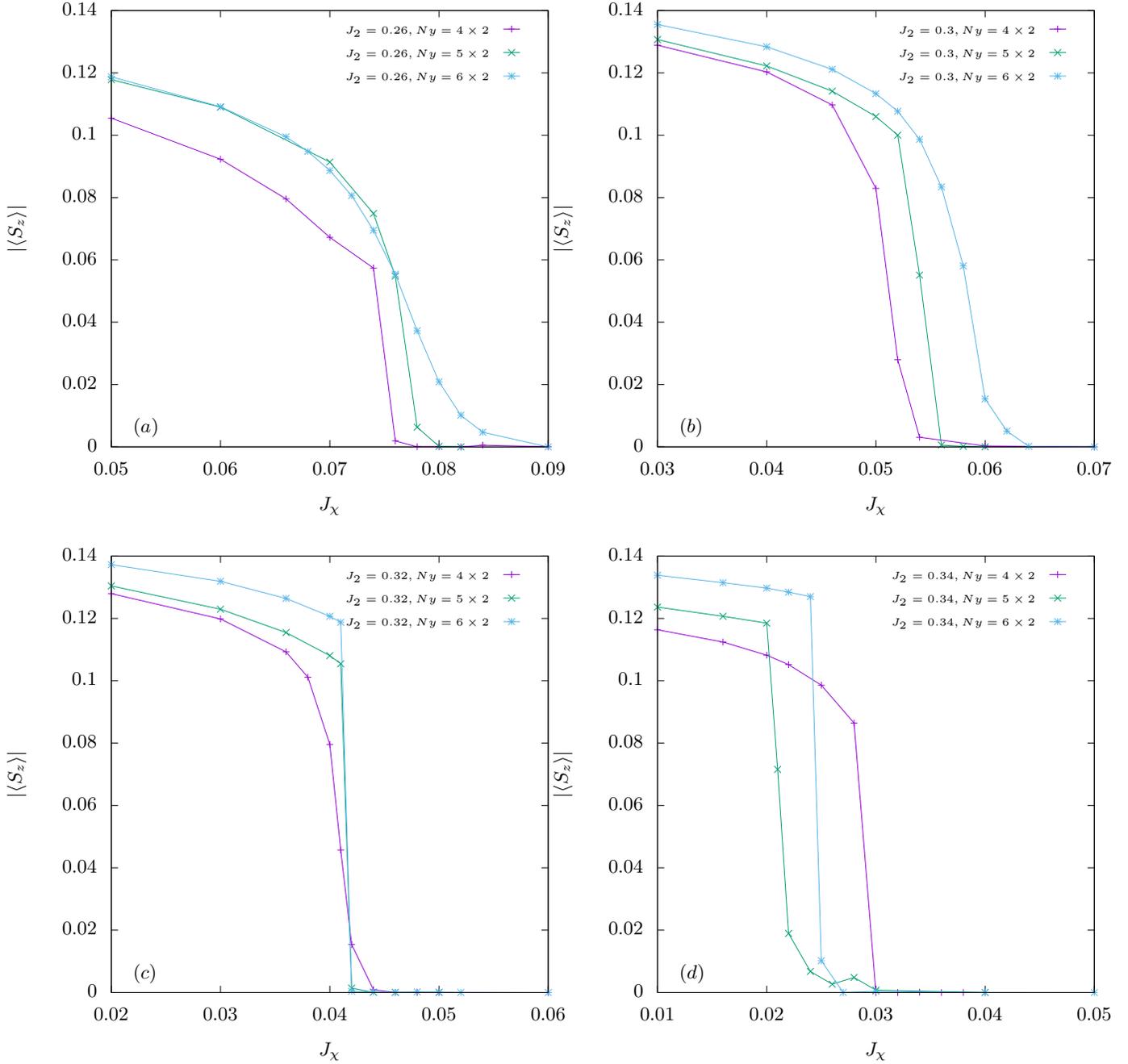}
\caption{\label{FigS4}(Color online)  The staggered magnetization at $J_{2}=0.26$ (a), $J_{2}=0.3$ (b), $J_{2}=0.32$ (c), and $J_{2}=0.34$ (d) for various sizes in the y-direction in the intermediate regime. The phase transition from the Ising antiferromagnetic state to the CSL is determined by the sudden drop of $\left | \left \langle S_{z} \right \rangle \right |$.}
\end{figure}

To further test the finite size effect we study various system sizes with infinite DMRG, because the system length is already in the thermodynamic limit in the x-direction. While the CSL is robust in various sizes as shown in Fig.\ref{FigS10}, the critical $J_{\chi }$ of the phase boundary between the Ising antiferromagnetic state and the CSL in the intermediate regime may vary slightly. Here we show some examples in Fig.\ref{FigS4}. The critical $J_{\chi }$ remains finite for all values of $J_{2}$ in the thermodynamic limit.

\begin{figure}
\centering
\input{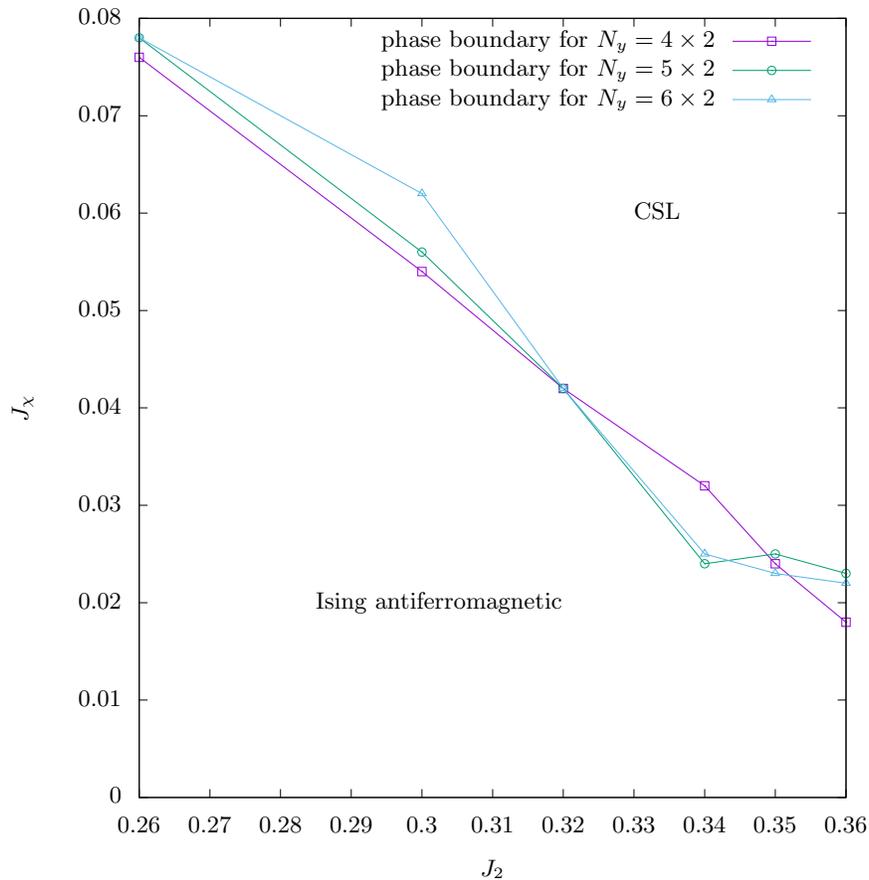}
\caption{\label{FigS5}(Color online)  The phase boundary for various sizes in the y-direction in the intermediate regime using infinite DMRG.}
\end{figure}

\begin{figure}
\centering
\input{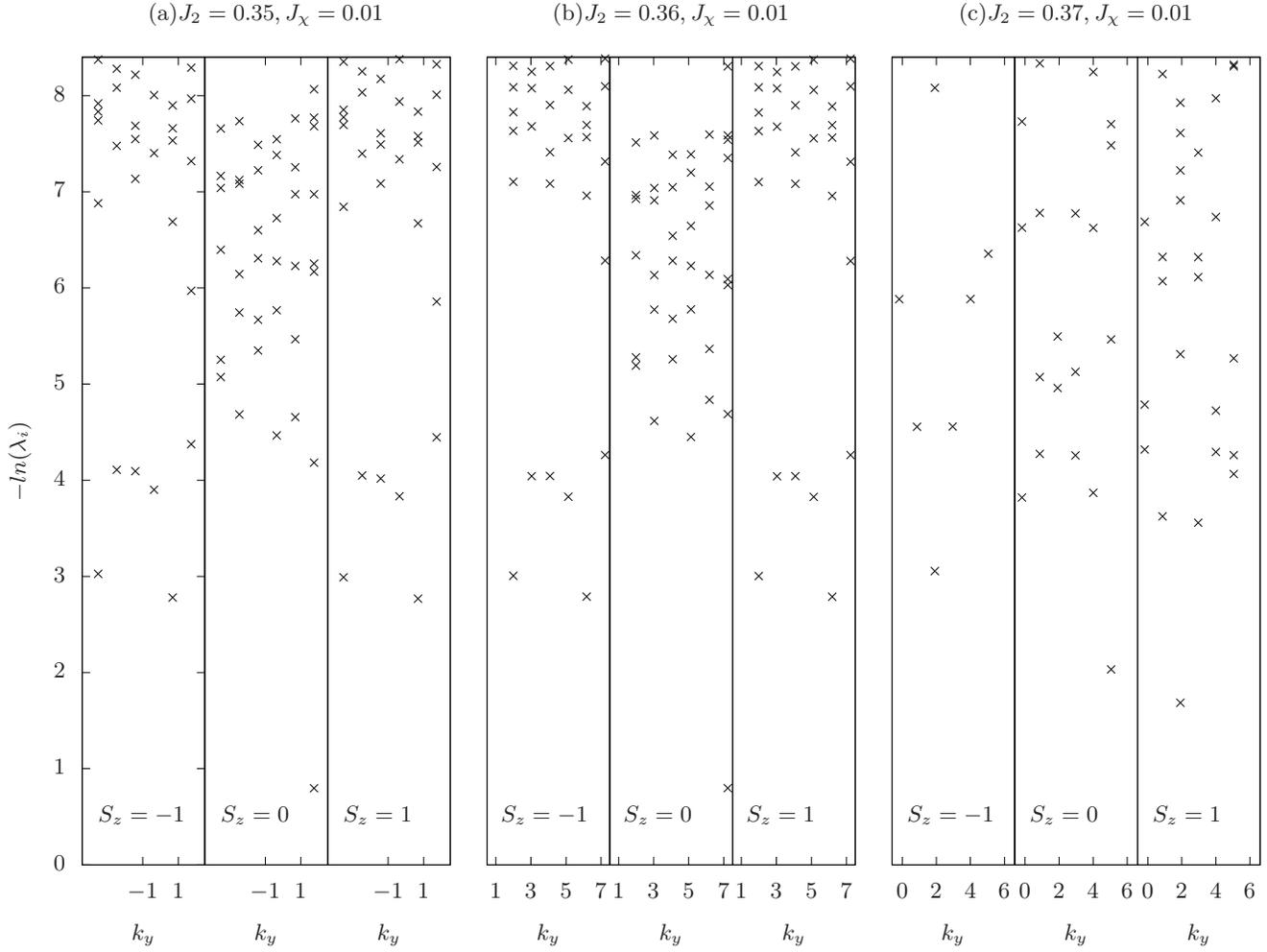}
\caption{\label{FigS6}(Color online)  The entanglement spectrum for $J_{\chi }$ bellow the critical $J_{\chi }$ of the phase boundary, which is in the magnetic regime. There is no quasi-degenerate pattern because of the additional low energy spectrums that mixed in.}
\end{figure}

The phase boundary for various system sizes between the CSL and the Ising antigerromagnetic phase in the intermediate regime is given in Fig.\ref{FigS5}. For $J_{2}\sim 0.36$ the staggered magnetization becomes too small and the phase boundary is determined by the entanglement spectrum. We have found that the phase boundary remains almost the same for various system sizes, and for $J_{\chi }$ smaller than the critical $J_{\chi }$ of the phase boundary, there is no quasi-degenerate eigenvalues in the entanglement spectrum, as some of the examples given in Fig.\ref{FigS6}.

\begin{figure}
\centering
\input{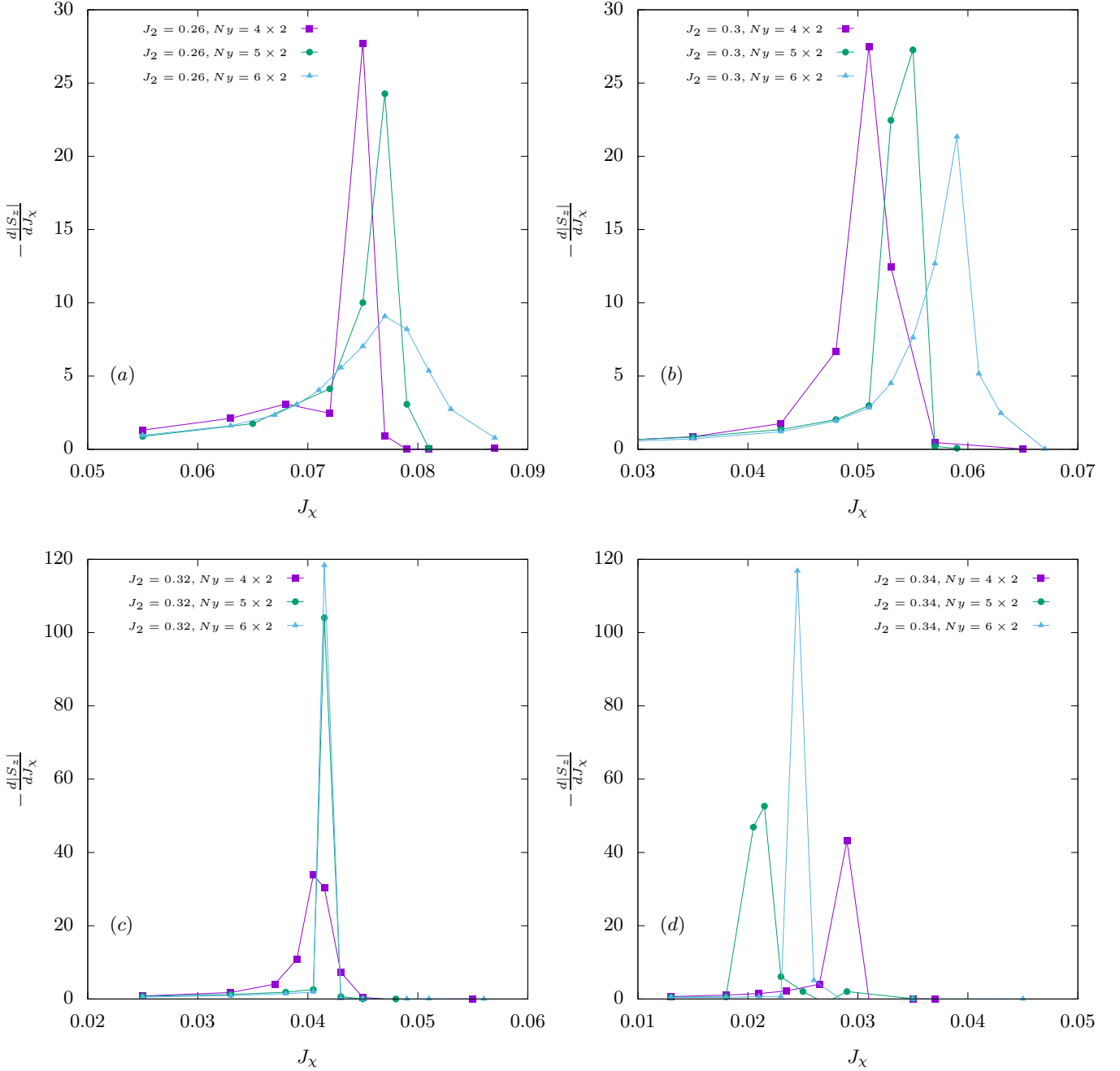}
\caption{\label{FigS8}(Color online)  The derivative of $\left | \left \langle S_{z} \right \rangle \right |$ with respect to $J_{\chi }$ at $J_{2}=0.26$ (a), $J_{2}=0.3$ (b), $J_{2}=0.32$ (c), and $J_{2}=0.34$ (d) for various sizes in the y-direction in the intermediate regime.}
\end{figure}

From Fig.\ref{FigS4} we can already see that the order parameter of the Ising antiferromagnetic state $\left | \left \langle S_{z} \right \rangle \right |$ decreases slower for larger system sizes when $J_{2}<0.3$, indicating a higher order phase transition. To further clearify we plot the $ -\frac{d|S_{z}|}{dJ_{\chi}} $ for different system sizes. As shown in Fig.\ref{FigS8}, the peak decreases as system size increase for $J_{2}<0.3$, and it increases as system size increase for $J_{2}>0.3$. Although larger system size with smaller truncation error may be needed to identify the transition type, the behavior of the peak is consistent with a higher order phase transition for $J_{2}<0.3$ and first order transition for $J_{2}>0.32$.

\section{The entanglement spectrum near the phase boundary}

\begin{figure}
\centering
\input{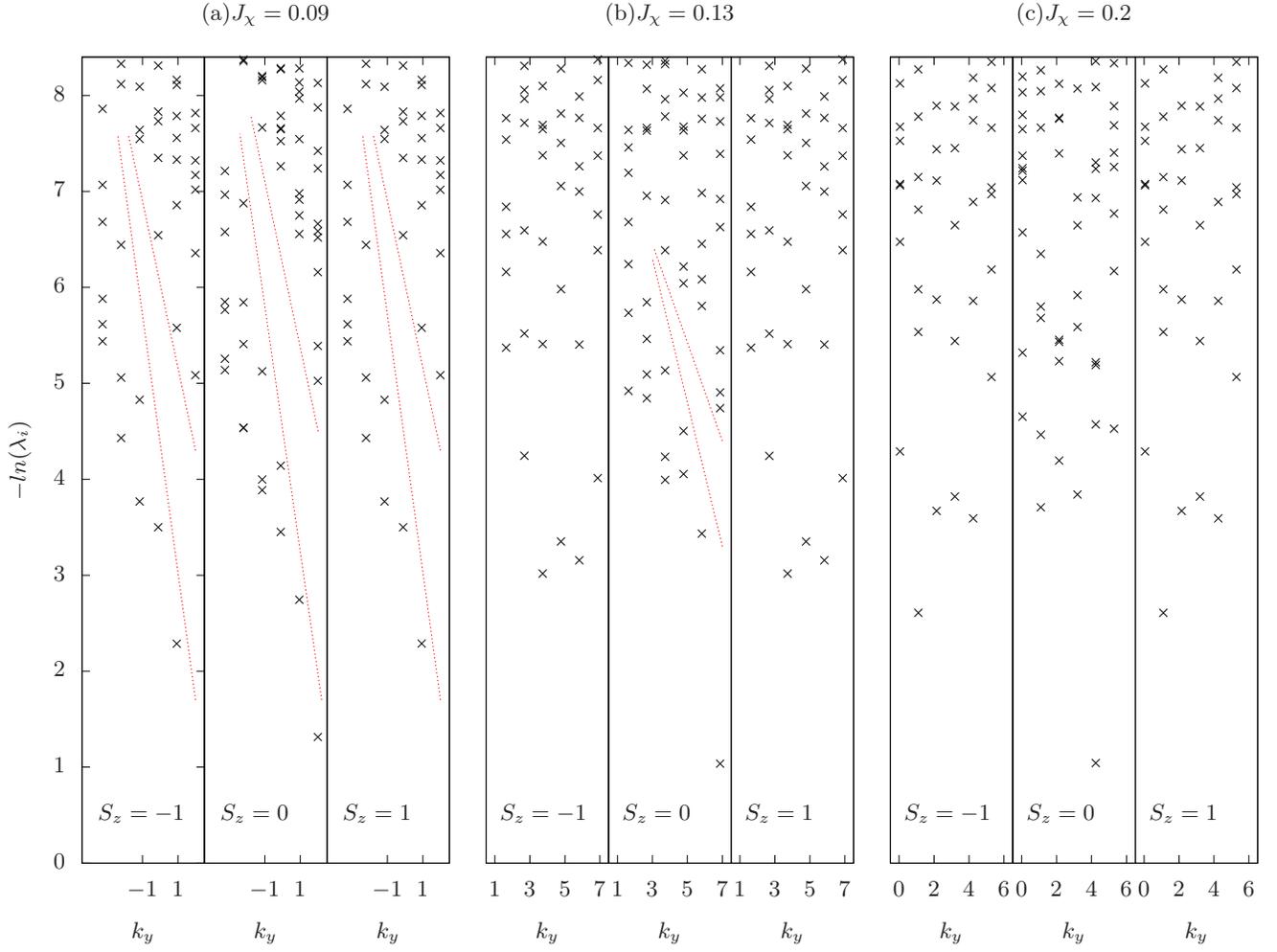}
\caption{\label{FigS7}(Color online) The entanglement spectrum in CSL at $J_{\chi }=0.09$ (Fig.\ref{FigS7}(a)), boundary at $J_{\chi }=0.13$ (Fig.\ref{FigS7}(b)), and the chiral spin state at $J_{\chi }=0.2$ (Fig.\ref{FigS7}(c)). All three plots are obtained at $J_{2}=0.26$. The dash red lines are guide to the eye.}
\end{figure}

The entanglement spectrum provides an effective way to determine the phase boundary between the CSL and the chiral spin state. As shown in Fig.\ref{FigS7}(a), the counting of the quasi-degenerate states in the CSL is very clear in every spin sector. In Fig.\ref{FigS7}(b) we can still identify the counting for $S=0$ sector near the phase boundary, but additional low-lying eigenstates have already mixed in the the $S=-1$ and 1 sector. As soon as the system enters the chiral spin state the counting disappears, as shown in Fig.\ref{FigS7}(c).

\section{The spin structure in chiral spin state}

\begin{figure}
\centering
\input{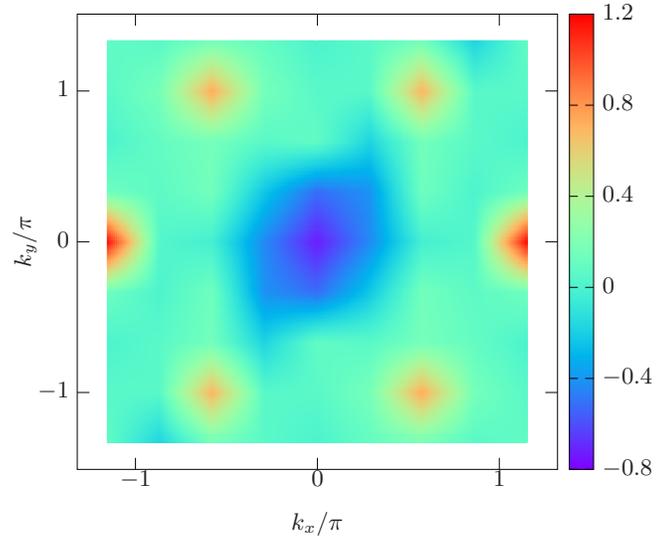}
\caption{\label{FigS9}(Color online) The spin structure in the chiral spin state at $J_{2}=0.2,J_{\chi }=0.27$, which peaks at the $M$ points in the second Brillouin zone.}
\end{figure}

The spin structure in the chiral spin state is given in Fig.\ref{FigS9}, which is calculated at $J_{2}=0.2,J_{\chi }=0.27$ using finite size cylinder of $L_{x}\times L_{y}=20\times 4\times 2$. There are 6 peaks at $M$ points, which resembles the spin structure of the tetrahedral phase in the extended Heisenberg model with three-spins chiral interactions on the honeycomb lattice\cite{hickey2017emergence}.

\twocolumngrid

\bibliography{Hc_XY}

\end{document}